
%
%
\input vanilla.sty
\input epsf.tex
\font\tenbf=cmbx10

 1
 1
 1
\font\ninebf=cmbx9
\font\ninerm=cmr9
\font\nineit=cmti9

\font\eightrm=cmr8
\font\eightit=cmti8

\TagsOnRight
\pageno=1
\def\firstfootline{\eightrm\hss\folio\hss}
\def\otherfootline{\eightrm\hss\folio\hss}
\footline={\ifnum\pageno=0\firstfootline\else\otherfootline\fi}
\catcode`@=11                   
\newinsert\footins
\def\norulefootnote#1{\let\@sf\empty
        \ifhmode\edef\@sf{\spacefactor\the\spacefactor}\/\fi
        {\eightrm #1}\@sf\vfootnote{\eightrm #1}}
\def\vfootnote#1{\insert\footins\bgroup
        \interlinepenalty\interfootnotelinepenalty
        \splittopskip\ht\strutbox
        \splitmaxdepth\dp\strutbox \floatingpenalty=20000
        \leftskip\z@skip \rightskip\z@skip \spaceskip\z@skip
\xspaceskip\z@skip
        \noindent{\eightrm #1}\footstrut\futurelet\next\fo@t}
\def\fo@t{\ifcat\bgroup\noexpand\next \let\next\f@@t
        \else\let\next\f@t\fi \next}
\def\f@@t{\bgroup\aftergroup\@foot\let\next}
\def\f@t#1{\eightrm #1\@foot}
\def\@foot{\strut\egroup}
\def\footstrut{\vbox to\splittopskip{}}
        \skip\footins=\bigskipamount
        \count\footins=1000
        \dimen\footins=8in
        \output{\plainoutput}
\def\plainoutput{\shipout\vbox{\makeheadline\pagebody\makefootline}%
        \advancepageno
        \ifnum\outputpenalty>-20000 \else\dosupereject\fi}
\def\pagebody{\vbox to\vsize{\boxmaxdepth\maxdepth \pagecontents}}
\def\makeheadline{\vbox to\z@{\vskip-22.5\p@
        \line{\vbox to8.5\p@{}\the\headline}\vss}\nointerlineskip}
\def\makefootline{\baselineskip24\p@\line{{\eightrm\the\footline}}}
\def\dosupereject{\ifnum\insertpenalties>0
        \line{}\kern-\topskip\nobreak\vfill\supereject\fi}
\def\pagecontents{\ifvoid\topins\else\unvbox\topins\fi
        \dimen@=\dp255 \unvbox255
        \ifvoid\footins\else \vskip\skip\footins \unvbox\footins\fi
        \ifr@ggedbottom \kern-\dimen@ \vfil \fi}

\hsize=5.0truein
\vsize=7.8truein
\baselineskip=10pt

\def\qed{\hbox{${\vcenter{\vbox{
    \hrule height 0.4pt\hbox{\vrule width 0.4pt height 6pt
    \kern5pt\vrule width 0.4pt}\hrule height 0.4pt}}}$}}
\def\vec#1{\underline{#1}}
\def\mat#1{\vec{\vec{#1}}}
\def\laplace{\Delta}
\def\grad{\nabla}
\def\div{\nabla\cdot}
\newcount\eqnnumber \eqnnumber=1
\def\num{\tag\the\eqnnumber \global\advance\eqnnumber by 1}
\font\eighttt=cmtt8
\def\setupverb{\parskip 0pt
               \eighttt
               \obeylines
               \obeyspaces }{\obeyspaces\global\let =\ }
\def\listing#1#2{\smallskip
                 \begingroup\setupverb { #2 } \endgroup
                 \centerline{\eightrm#1}
                 \medskip\hrule\medskip\medskip\noindent\ignorespaces}
\newcount\sectionnumber \sectionnumber=1
\def\section#1{%
\vglue 12pt
\line{\tenbf \the\sectionnumber . #1\hfil} \global\advance\sectionnumber by 1
\vglue 5pt
\noindent\ignorespaces}
%
%
%
%
\vglue 5pc
\baselineskip=13pt
\centerline{\bf CELLULAR AUTOMATON}
\centerline{\bf FOR THE FRACTURE OF ELASTIC MEDIA}
%
%
\vglue 24pt
\centerline{\eightrm PETER OSSADNIK}
\baselineskip=12pt
\centerline{\eightit HLRZ, KFA J\"ulich}
\baselineskip=10pt
\centerline{\eightit Postfach 1913, D-5170 J\"ulich, Germany}
\vglue 20pt
\baselineskip=10pt
%
%
\vglue 16pt
{\rightskip=1.5pc
 \leftskip=1.5pc
 \eightrm\baselineskip=10pt\parindent=1pc\noindent
We study numerically the growth of a crack in an elastic medium under
the influence of a travelling shockwave. We describe the implementation of
a fast algorithm which is perfectly suited for a data parallel computer.
Using large scale simulations
on the Connection Machine we generate cracks with more than 10000 sites on
a $\scriptstyle 1024 \times 1024$ lattice. We show that the resulting patterns
are
fractal with a fractal dimension that depends on the chosen breaking
criterion and varies between $\scriptstyle 1.$ and $\scriptstyle 2.$
\vglue 12pt}
%
%
\baselineskip=13pt
\section{Introduction}
How does a solid body break under an externally applied load? This is an
important question for researchers in many fields and has strong
implications on e.g. materials science, engineering or geophysics. It
has been studied for a long time quite extensively using numerical,
experimental and analytical methods.$^1$

If one considers the solid to be a linear elastic medium the formulation
of the initial question can be made in terms of the Lam\'e equation
$$(1-2\nu)\laplace\vec u + \grad( \div \vec u) = 0 \num$$
which describes the displacement $\vec u$ of a small volume element in
an elastic material from its equilibrium position. The Poisson ratio
$\nu$ is a material dependent parameter which has the following meaning: If one
applies a uniaxial force to an elastic bar of length $L$ and cross
section $W\times W$, it will not
only change its length by $\Delta L$ in the direction of the force, but
it will also change its width by $\Delta W$ in the direction
perpendicular to the force. The Poisson ratio then defines the relative
amount of change
$$\nu=-{{\Delta W/W}\over{\Delta L/L}}.\num$$
Due to thermodynamical reasons $\nu$ is bounded between $-1\leq\nu<0.5$.
Concrete for instance has a Poisson ratio $\nu\approx 0.2$.

A crack in such a system can be described as an additional force free
surface which thus obeys the boundary condition
$$\mat\sigma\cdot\vec n=0\num$$
where $\mat\sigma$ is the stress tensor and $\vec n$ is a surface
vector. The crack grows if the stress parallel to the crack surface
$\sigma_\parallel$
is larger than a certain material dependent critical stress $\sigma_C$.
Since no first principle law for the normal growth velocity is known one
assumes the general behavior
$$v_n\propto (\sigma_\parallel-\sigma_C)^\eta \num$$
where $\eta$ is often simply set to unity.
The equations (1), (3) and (4) formulate the growth of a
crack as a moving boundary problem which is far more difficult to solve
than the Lam\'e equation itself. Yet, there is one important property
of real material missing, which is ``disorder''. Microscopically
disorder means deviations from the perfect crystal structure of the
elastic material due to
vacancies, dislocations or grain boundaries. But macroscopically these
imperfections are simply reduced to spatial randomness of the material
properties like Poisson ratio and breaking threshold.

In numerical simulations for crack growth
one often discretizes the elastic medium on a lattice,
for instance with a finite element scheme.$^2$ According to the chosen
boundary conditions and the externally applied load one relaxes the
system to equilibrium. Then one
picks one or several bonds according to a given rule -- for instance the
bond with the highest load -- and breaks it. This defines a new boundary
and therefore one has to resolve the whole problem again. This procedure is
then repeated until a crack of desired size is grown. The disorder is often
put into the simulations by considering lattice parameters (like bond strength
and breaking threshold) that vary from site to site.
Since the system has enough time to relax to full equilibrium before
the crack can grow, such a procedure is only capable of
describing quasi static  processes.

\headline={\ifodd\pageno\rightheadline \else \leftheadline\fi}
\def\rightheadline{\hfil\eightit Cellular Automaton for the
     Fracture of Elastic Media\quad\eightrm\folio}
\def\leftheadline{\eightrm\folio\quad \eightit P. Ossadnik\hfil}
\voffset=2\baselineskip
\nopagenumbers

On the other hand there are phenomena --- like explosions, shattering
of glass or shock waves -- which should not be treated in a static
approximation and in which
the system is not able to relax to equilibrium before breaking a bond,
but in which the time to relax the system is comparable to or larger
than the time one needs to propagate the crack. Such a situation usually
results in cracks with many sidebranches growing behind the shock front
since the internal energy cannot be dissipated fast enough.$^3$
In the following we are going to study numerically such a process in
which one obviously has to take into account the dynamical behavior of
the elastic medium.

The organization of this paper is thus as follows: in Sect.~2 we
introduce the general model, whose implementation on the Connection Machine
is described in Sect.~3. In Sect.~4 we describe the details of the
simulation and in Sect.~5 we present some results.
\section{The Model}
As a model for the elastic material we consider a triangular network of
Hookean springs connecting points of mass $m$. A triangular network is
necessary for the simulation since a simple quadratic lattice does not have
any shear modulus and thus can be deformed arbitrarily under shear load.
This model is known as
``central force model''$^2$ since the Hookean springs are isotropic.
The boundary sites of this network are kept fixed in space.
On each site there are two continuous degrees
of freedom, which are the coordinates of the displacement
 $u_{x}$ and $u_{y}$ of this site
from its equilibrium position $\vec r_0$.
Since we want to study the dynamical behavior of this network we have to
determine the time evolution of the displacements which is governed on
each lattice site by Newton's equation.
Following a suggestion of Chopard$^4$ we use a
discrete time Hamilton formalism to express the dynamic behavior.
The Hamiltonian of this system is given by
$$\eqalign{H
\left( {\vec p_{1}
\ldots \vec p_{N},\vec r_{1}
\ldots \vec r_{N}}
\right) &=
\sum \limits _{i=1}^{N}{{\vec p_{i}^{2}} \over {2mi}}+{{1} \over {2}}
\sum \limits _{i,j=1}^{N}U_{ij}
\left( {\vec r_{i},\vec r_{j}}
\right) \cr}
\num$$
where $\vec p_i$ and $\vec r_i$ are the momentum and position of site $i$,
and $U_{ij}$ is the interaction energie between two lattice sites $i$
and $j$. $U_{ij}$ is nonzero only between nearest neighbor sites and since we
use
Hookean springs it is simply a harmonic potential
$$
\eqalign{
U_{ij}&=
{{k_{ij}} \over {2}}
\left( {
\left\vert {
{\vec r_{i}-\vec r_{j}}}
\right\vert -a}
\right) ^{2}
=
{{k_{ij}} \over {2}}
\left( {
\left\vert {
{\vec u_{i}+\vec r_{0i}-\vec u_{j}-\vec r_{0j}}}
\right\vert -a}
\right) ^{2}
\cr
&=
{{k_{ij}} \over {2}}
\left( {
\left\vert {
{\vec u_{i}-\vec u_{j}+\vec{dr}_{ij}}}
\right\vert -a}
\right) ^{2}
\cr
}\num$$
where $k_{ij}$ is the coupling constant between the sites $i$ and $j$.
${\vec{dr}_{ij}}={\vec r_{0i}-\vec r_{0j}}$
is the vector between their equilibrium positions
and $a$ is the equilibrium length of the connecting spring (please note,
that ${\vec{dr}_{ij}}$ does not mean differentials). In our
simulation we are going to set $a=0$ while we keep $\vec{dr}_{ij}=1$.
This corresponds to the case of a prestretched network and can be
compared for instance with the skin of a drum.
The discretized versions of Hamilton's equations are
$$\eqalign{
\vec r_{i}\left( {t+1}
\right) -\vec r_{i}
\left( {t}
\right) &={{
\partial H} \over {
\partial \vec p_{i}}}={{\vec p_{i}
\left( {t}
\right) } \over {m_{i}}}\cr \vec p_{i}
\left( {t}
\right) -\vec p_{i}
\left( {t-1}
\right) &=-{{
\partial H} \over {
\partial \vec r_{i}}}\cr }\num
$$
where we use vector derivatives as symbols for the corresponding gradients.
The time evolution of the displacement field can finally be written
as
$$\eqalign{
\vec u_{i}
\left( {t+1}
\right) -2\vec u_{i}
\left( {t}
\right) + \vec u_{i}
\left( {t-1}
\right) &=-{{1} \over {m_{i}}}{{
\partial H} \over {
\partial \vec u_{i}(t)}}\cr
&=\sum_{j={NN}(i)}k_{ij}\cdot \vec \delta_{ij}
\cdot(1-{a\over \left|\vec \delta_{ij}\right|})\cr} \num
$$
where $\vec\delta_{ij}=\vec u_i(t)-\vec u_j(t)+\vec{dr}_{ij}$.
This equation (8) defines the  updating rule, which relates the
displacements at time $t+1$ to the displacements at times $t$ and $t-1$.
Although the chosen dynamics seems to be rather
crude as compared to molecular dynamics, one can actually show
that it conserves the total momentum and some kind of total ``energy''.
Moreover we show that it conserves the essential features we need for the
 generation of cracks: A wave packet that is imposed on the lattice will travel
with only minor changes in shape with a definite velocity through the lattice.
\section{The Growth Model using Fortran 90}
Since the same updating rule (8) is applied to all lattice sites at each
time step and on the other hand the topology of the underlying lattice is,
in contrast to molecular dynamics,
not changed due to rearrangements of particles
single instruction multiple data
 (SIMD) machine like the CM is the appropriate computer
architecture for this problem: Each lattice site is mapped onto one
virtual processor and all sites are updated in parallel.
The only inter processor communication is required for the calculation
of the right hand side of eqn.~(8). But
because the triangular lattice structure can be mapped onto a square
lattice with next nearest neighbor interactions into one direction, a
nearest neighbor grid communication using {\tt CSHIFTS} is sufficient
and no general router communication is required, which makes this lattice
model fast and efficient.

In the actual implementation of the program we chose the following data
layout for the main variables:
Each processor has to store the displacements at time $t$ and
$t-1$, $\tt UT$ and $\tt UTM1$, which are represented as complex
numbers. Since we intend to simulate a
``disordered'' system, we assign to each spring a different, randomly chosen
coupling constant. Therefore we keep on each
lattice site the six couplings $\tt K$ to all neighbors.
By doing this we waste
some memory space because each $k_{ij}$ is stored twice -- on site $i$ and on
site $j$ -- but we save computer time by avoiding unnecessary
communication. The data layout for the main variables is thus as follows
\begingroup\setupverb
      COMPLEX , ARRAY (NXY, NXY)    :: UT, UTM1
      REAL    , ARRAY (Z, NXY, NXY) :: K
CMF\$  LAYOUT UT(:NEWS,:NEWS)
CMF\$  LAYOUT UTM1(:NEWS,:NEWS)
CMF\$  LAYOUT K(:SERIAL,:NEWS,:NEWS)
\endgroup
\listing{Listing 1.}{}
The {\tt LAYOUT} directive for {\tt K}
is used to group the couplings of one site as
a serial dimension
onto one processor. It is necessary, because otherwise the compiler would
spread the couplings over all virtual processors which results in
unnecessary communication for the force calculation.
Using this data layout a single step of the updating
rule (8) is programmed in a straightforward manner in CM Fortran
\begingroup\setupverb
C        GET VECTORS TO NEIGHBORS
C        COMPUTE NEW PARTICLE POSITION AND MOMENTUM
C        NN is a help field to store r\_i-r\_j. It has the same layout as
C        ut and utm1.
         NN(1,:,:) = CSHIFT(UT,1,1)
         NN(2,:,:) = CSHIFT(UT,2,1)
         NN(4,:,:) = CSHIFT(UT,1,-1)           !get displacements
         NN(5,:,:) = CSHIFT(UT,2,-1)           !of nearest neighbors
         NN(3,:,:) = CSHIFT(NN(4,:,:),2,1)
         NN(6,:,:) = CSHIFT(NN(1,:,:),2,-1)

         NN(1,:,:) = NN(1,:,:)-UT+DR\_1
         NN(2,:,:) = NN(2,:,:)-UT+DR\_2
         NN(3,:,:) = NN(3,:,:)-UT+DR\_3
         NN(4,:,:) = NN(4,:,:)-UT+DR\_4         !calculate difference vector
         NN(5,:,:) = NN(5,:,:)-UT+DR\_5
         NN(6,:,:) = NN(6,:,:)-UT+DR\_6

         UTM1 = SUM(NN*(K*(1.-A1/ABS(NN))),DIM=1)/M + 2.*UT - UTM1
\endgroup
\listing{Listing 2.}{}
The {\tt CSHIFT} commands are used to communicate the displacements
between ``neighboring'' processors. By carefully reusing already shifted
data it is of course possible to get the displacements from the six
nearest neighbors on the triangular lattice with only six {\tt
CSHIFTS}. Since we are not using full next nearest
neighbor communication the use of stencil operations does not seem
useful at this point.
The global {\tt SUM} along the first
dimension computes the total force exerted onto each site by its
neighbors. Because the first dimensions of the coupling constant array
{\tt k} and the vectors to the nearest neighbors {\tt NN} are laid out
onto the same processor as a {\tt :serial} dimension,  the computation
of the {\tt SUM} does not require any communication.

Unfortunately this simple formulation does not give optimal performance.
The compiler allocates and deallocates unnecessary temporary fields
and even generates general {\tt CM\_send} router communication, which in
fact uses 43\% of the total CPU time!
To obtain a much better performance
we coded the code fragment shown in Listing 2 completely in PARIS (PARallel
Instruction Set).
This allows to fully control the
memory allocation, the communication and to make efficient use of
pipelined commands like {\tt CM\_f\_sub\_const\_mult\_always}. Now, most
of the CPU time -- 38\% -- is used for the calculation of the distance between
neighboring lattice sites $|\vec r_i-\vec r_j|$ which is coded as {\tt
CM\_f\_c\_abs\_2}. Now, the NEWS communication part is negligible and sums up
to 5.8\% of the total CPU time.

These improvements result in a speedup of a big factor of five as
compared to the straightforward implementation. One obtains on
8K processors of the previously described CM2 an update rate of
1.1 millions of updates per second (MUPS) and a speed of 110 MFlops.
As a comparison, one can obtain with typical spin cellular automata using
multispin coding techniques more than 1000 MUPS on one
processor of a NEC-SX3$^5$ and on a CM2-16K a Q2R cellular automaton runs at
1600 MUPS.$^6$ For a full
molecular dynamics simulations on an CM200-8K Hedman and Laaksonen achieve
about
0.2 MUPS$^7$
and for MD simulations on a CM2-16K Mel`\v cuk et.~al.
obtained an update rate of 4.5 KUPS.$^8$
\section{The simulation}
All simulations are performed on a $ 1024\times 1024 $ lattice and we use
between 3000 and 5000 timesteps.
At the beginning of the simulation the couplings are chosen at random
out of a uniform distribution with mean value $k_0 = 0.005 $ and a width
of typically 50\%. To initiate the crack growth we break all bonds between site
$\vec r_0$ and its neighbors.
Afterwards an initial pulse is imposed on the center of the lattice:
If ${\vec r_{0}}$ is the central site in the lattice, then
its nearest neighbors $\vec r_1\ldots\vec r_6$ are displaced radially outward
for 100 lattice units while keeping all other sites fixed. This displacement
seems very large at first sight, but since we use a harmonic potential between
the sites the actual size of the initial displacement is not relevant. We only
wanted to make sure that this perturbation is much larger than the ``thermal''
motion
unduced by chosing random coupling constants.
At time $t=0$ all sites are released and the system is free to evolve.
After every other time step one looks for the bond with the largest
elongation $l_{m}$. Then one determines all bonds whose elongations $l$
are larger than $\alpha \cdot l_{m}$ -- where $\alpha$ is an adjustable
parameter -- and which lie on the surface of the already existing crack.
All those bonds are broken by setting the corresponding
coupling constant $k_{ij}$ to zero. Here one has to notice that this
breaking rule does not require any communication and can be performed
completely in parallel since both sites $i$ and $j$ which are connected
by such a bond will clear their own copy of $k_{ij}$.

Thus we consider a relative breaking threshold rather
than an absolute one, which has the following reason:
In an absolute breaking threshold one would break all bonds whose
elongation is larger than some fixed critical length $l_c$. On the other
hand the amplitude of the outgoing wave packet is decreasing with the
distance from the center.
So, when the wave packet has initially an amplitude that is larger than
the threshold, the outgoing wave will break all bonds it reaches until
its amplitude has dropped below the threshold and from then on no
further bond will be broken. Thus, one only creates a structureless isotropic
hole in the center of the system.

A somewhat similar model has been
studied by Louis et.~al.$^9$  They try to solve a quasi static problem,
but perform only a
few relaxation steps to find the equilibrium state of the network.
However, since they use an overrelaxation scheme the relaxation of their system
towards equilibrium has another dynamical interpretaion than the
iterative method I use.
Another similarity is given in the breaking rules.
Louis et.~al. pick bonds that are to be broken with a probability
that is proportional to the bond length
$P_{ij}\propto|\vec r_i-\vec r_j|$.
\section{Results}
To demonstrate that the dynamics (8) produces reasonable results we
first consider the case of a smooth wave packet travelling through a
system without disorder and without breaking.
Therefore we applied not a singular pulse to the network but
rather a smooth wave packet. The initial radial displacements of the
central sites are
$$\vec u(t) = \vec
u_0\cdot\left(1-cos\left(2\pi\cdot{t\over\tau}\right)\right)\num$$
for $0\leq t\leq \tau$ and $\vec u(t)=0$ for $t>\tau$ while all other sites
are free to move. The period is chosen to be $\tau=100$.
\midinsert
\def\epsfsize#1#2{\epsfxsize}\centerline{\epsfbox{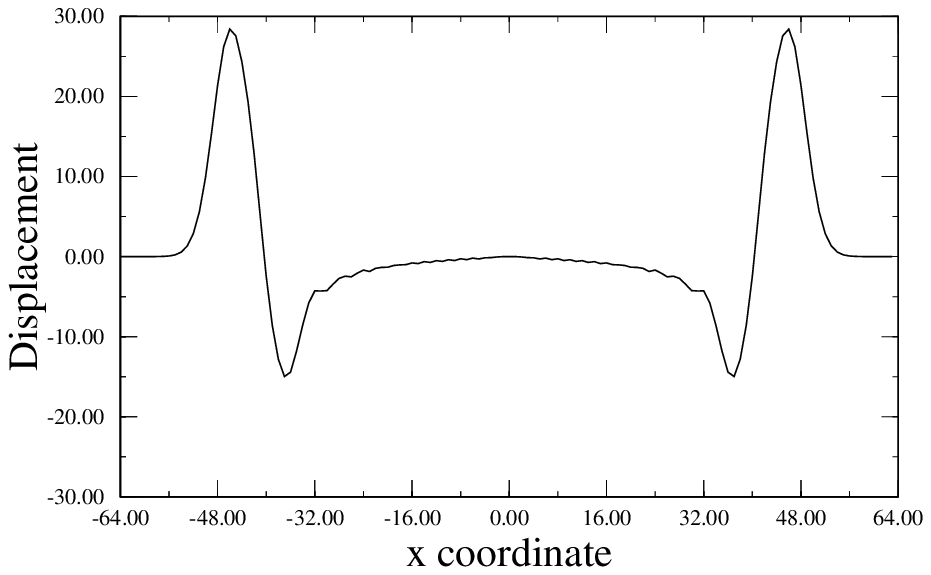}}
\centerline{\eightrm Fig.~1.\,\,\,Cut through a shockwave after 200 timesteps.}
\endinsert\noindent\ignorespaces
Fig.~(1) shows a cut along the x axis through this wave packet at time $t=200$.
One indeed recovers the original pulse plus a further minimum
which is due to the fact that we keep the displacements of the central sites
for $t>\tau$ fixed at zero. Because of their inertia the neighboring sites
keep vibrating which results in the second minimum.
If one measures the velocity of the maximum of the wave packet one
obtains a value which agrees
with the analytical expression for the group velocity of a
circular wave with frequency $\omega = 2\pi / \tau$ on a triangular lattice.

After having confirmed that the travelling wave shows reasonable
behavior we restrict ourselves again to the case of a singular pulse,
which corresponds to the case $\tau = 2$.
\midinsert
\def\epsfsize#1#2{0.25\hsize}
\centerline{\eightrm Fig.~2.\,\,\,Crack patterns generated for (from left
to right)
a) $\alpha=1$, b) $\alpha=0.98$ and c) $\alpha=0.95$.}
\endinsert\noindent\ignorespaces
In figs (2.a-2.c)
we show examples of cracks which were
generated for different breaking thresholds: $\alpha=1$ (10934 sites)
-- which means that only the bond with the
largest elongation is broken -- $\alpha=0.98$ (16038 sites) and $\alpha=0.95$
(35837 sites). The number in brackets are the number of sites that are
``connected''
 by broken bonds.
All cracks show a starlike and fractal structure.
For decreasing $\alpha$ they become more and more ramified. This
is easily understood since with decreasing $\alpha$ many more bonds are
eligible to be broken. The delta-like excitation we impose initially on
the lattice is highly non-periodic and therefore produces waves with
many different frequencies, althoug lower amplitude. For small enough
$\alpha$ many of these waves can contribute to the growth of the crack.
But one also has to take into account another effect.
Since the lattice is prestretched
each bond that is broken is a source for another spherical
wave travelling away from this point. For small enough $\alpha$ also
those waves can break bonds and therefore can lead to an avalanche-type
growth of the crack. Thus, one can distinguish two different regimes:
For $\alpha$ close to unity the crack grows mainly at the tips at the outer
branches which coincide with the front of the shockwave. The
sidebranches behind the shockfront remain inactive. For smaller
$\alpha$ also the tips behind the shockfront continue to grow and split
and eventually the crack becomes space filling.

Another fact to be noticed is that for decreasing $\alpha$ the lattice
structure becomes more and more dominant and eventually the cracks grow
into a structure with sixfold symmetry.

To be more quantitative, we study the dependence of the number of broken
bonds $N$ on the radius of the cluster, which is measured in terms of the
radius of gyration $R_G$. $R_G$  describes the average distance of all
broken bonds from the center of mass $\vec r_{CM}$ of the crack
$$ R_G = \sqrt{\sum_{i=1}^{N} \left(\vec r_i-\vec r_{CM}\right)^2}.\num$$
\midinsert
\def\epsfsize#1#2{\epsfxsize}\centerline{\epsfbox{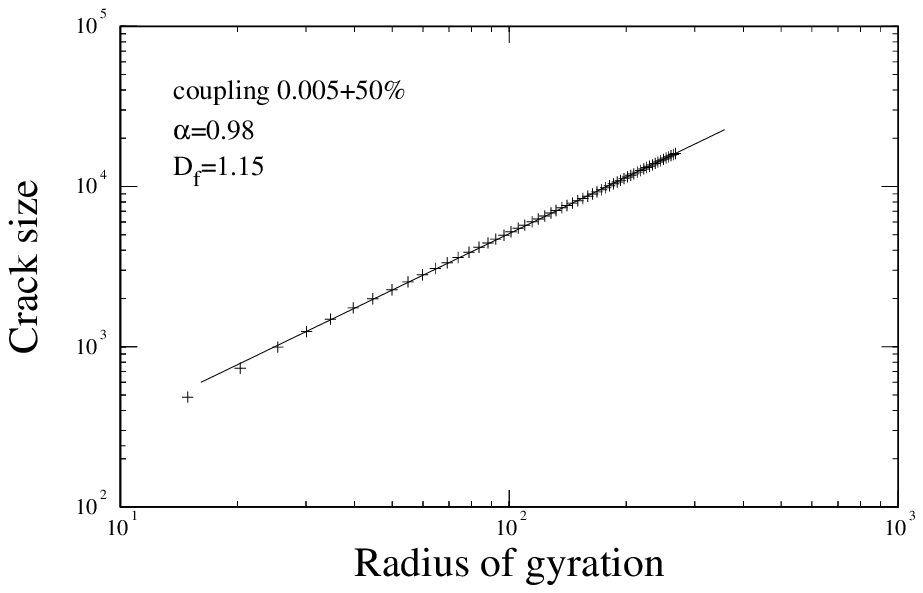}}
\centerline{\eightrm Fig.~3.\,\,\,Scaling of the number of broken bonds
with the radius of the crack.}
\endinsert\noindent\ignorespaces
In fig.~(3) we show the typical behavior of the number of broken
bonds. As an example we show data for $\alpha=0.98$ in which we averaged
over four cracks.
\midinsert
\def\epsfsize#1#2{\epsfxsize}\centerline{\epsfbox{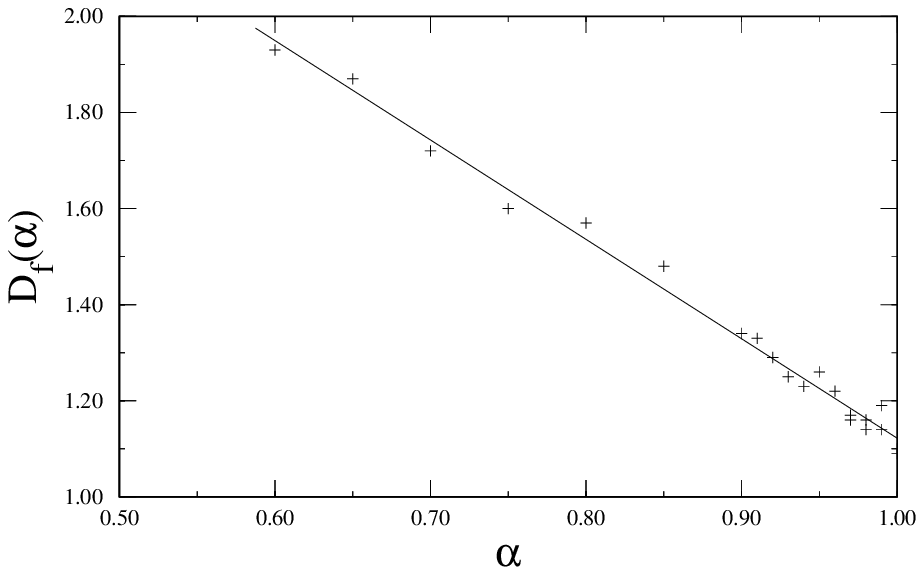}}
\centerline{\eightrm Fig.~4.\,\,\,Dependence of fractal dimension on the
breaking threshold.}
\endinsert\noindent\ignorespaces
One finds that this number has a power law dependence on the
radius of gyration
$$N\propto R_G^{D_f}\num$$
and for this specific example we find a fractal dimension $D_f = 1.15$.
As indicated above this exponent varies with varying $\alpha$.
So, in fig.~(4)
we show all exponents $D_f$ for different $\alpha$.

Also in this
plot the exponents $D_f(\alpha)$ were obtained by averaging over four
independent realizations. One obtains a linear dependence
$$D_f(\alpha) = -(2.06\pm 0.05)\cdot\alpha+(3.19\pm 0.05).\num$$
Thus, for decreasing $\alpha$ one approaches a space filling structure
which one reaches for $\alpha\approx 0.58$. However, it could be
possible that for even larger clusters the asymptotic behavior changes
and one crosses over into other exponents $D_f$.
\section{Conclusions}
We have described the implementation and results of a discrete time
simulation for the growth of large cracks on a triangular network.
We use a central force model and study the dynamical
behavior of crack growth instead of studying the slow growth modes.
Using a simplified dynamics, which anyway reproduces the essential features
for the crack production, we are able to grow cracks with more than
10000 sites on a $1024\times 1024$ lattice.
We obtain fractal growth patterns with dimensions
almost in the whole range between 1.1 and 2.0.
The dimensionality of the
cracks is mainly governed by the breaking threshold $\alpha$ and one
finds a linear dependence of $D_f$ on $\alpha$.
\vglue 12pt\line{\tenbf References\hfil}\vglue 5pt
\ninerm\baselineskip=11pt\frenchspacing
\item{1.} For recent reviews see
{\nineit Statistical Models for the Fracture of Disordered Media},  eds.
H.~J.~Herrmann, and S.~Roux (North-Holland, Amsterdam, 1990); {\nineit
Disorder and Fracture --- Proc. NATO ASI}, eds. J.~C.~Charmet, S.~Roux,
and E.~Guyon, (Plenum Press, New York, 1990); H.~J.~Herrmann, in
{\nineit Fractals and Disordered Systems}, eds. A.~Bunde, and S.~Havlin,
(Springer, 1990).
\item{2.} E.~Schlangen, J.~G.~M.~van~Mier, in {\nineit Proc.~Int.~EPRI.~
Conf. on Dam Fracture}, eds. V.~E.~Saouma et. al., 1991;
S.~Feng, P.~N.~Sen, {\nineit Phys.~Rev.~Let.} {\ninebf 52}, 216(1984);
S.~Roux, E.~Guyon, {\nineit J.~Physique.~Let.} {\ninebf 46}, L999(1985)
\item{3.} E.~H.~Yoffe, {\nineit Phil.~Mag.} {\ninebf 42}, 739(1951),
Lord Rayleigh, {\nineit Proc.~Lond.~Math.~Soc.} {\ninebf 17}, 4(1885)
\item{4.} B.~Chopard, {\nineit J.~Phys.~A} {\ninebf 23}, 1671(1990);
N.~H.~Margolus, Ph.~D. Thesis, MIT, 1987.
\item{5.} R.~W.~Gerling, D.~Stauffer, {\nineit Int.~J.~Mod.~Phys.~C}
{\ninebf 2}, 799(1991)
\item{6.} S.~C.~Glotzer, D.~Stauffer, S.~Sastry, {\nineit Physica A}
{\ninebf 164}, 1(1990)
\item{7.} F.~Hedman, A.~Laaksonen, in {\nineit Large Scale Computations
in Quantum Chemistry and Physics --- Proc.~Namur~SCF~Conf.},
eds.~J.~M.~Andr\'e et. al., (special issue of Int.~J.~Quant.~Chem.,
1992), to appear
\item{8.} A.~Mel'\v cuk, R.~Giles, and H.~Gould, {\nineit Computers in Physics}
{\ninebf May/June}, 311(1991).
\item{9.} E.~Louis, and F.~Guinea, {\nineit Europhys.~Let.} {\ninebf 3},
871(1987);
P.~Meakin, G.~Li, L.~M.~Sander, E.~Louis, and F.~Guinea, {\nineit J.~Phys.~A}
{\ninebf 22}, 1393(1989)

\bye